\input{aipcheck}
\documentclass[
    ,final            
  ]
  {aipproc}

\layoutstyle{6x9}

\begin{document}
\title{Gamma-rays from the compact colliding wind region in Cyg~OB2~\#5}

\keywords{stars: binaries; stars: individual: Cyg OB2 \#5; radiation mechanism: non-thermal; gamma-rays: theory}

\author{A. T. Araudo}{
address={Centro de Radioastronom{\'{\i}}a y Astrof{\'{\i}}sica, Universidad Nacional 
Aut\'onoma de M\'exico,  A.P. 3-72 (Xangari), 
58089 Morelia, Michoac\'an, M\'exico}
}
\author{G. N. Ortiz-Le\'on}{}
\author{L. F. Rodr{\'{\i}}guez}{}

\begin{abstract}
In this contribution we model the non-thermal emission (from radio to 
$\gamma$-rays) produced in the compact (and recently detected) colliding wind 
region in the multiple stellar system Cyg~OB2~\#5. 
We focus our study on the detectability of the produced $\gamma$-rays.
\end{abstract}

\maketitle


\section{Introduction}

The collision between stellar winds of massive stars is a  promising 
scenario to produce non-thermal emission.
In  early type binaries, 
a strong shock is created  between the stars, where the wind ram pressures 
are equated.  In those shocks particles can be 
accelerated up to relativistic energies via the Fermi-I acceleration mechanism,
and produce non-thermal emission. In particular, $\gamma$-rays can be
produced by the interaction of the accelerated particles 
with the  photon and matter fields provided by the stellar radiation and wind, 
respectively.   

Non-thermal radio emission from colliding wind regions (CWRs) 
has been firmly detected from many systems and they are putative 
candidates to be $\gamma$-ray sources since many years ago.  
In the past,
only marginal correlations between CWRs and unidentified $\gamma$-ray sources 
(of e.g. EGRET catalog) has been found. However, the recent detection of 
Eta-Car by \emph{Fermi} [1] and \emph{Agile} [2] shows that the $\gamma$-ray 
emission of CWRs is not only a theoretical prediction. 
In this work we study other promising $\gamma$-ray source to be detected by 
\emph{Fermi}: the peculiar system Cyg~OB2~\#5.

\subsection{The source Cyg OB2 \#5}

Cyg~OB2~\#5 is a radio-bright early-type
multiple system located into the big stellar association  Cyg~OB2, at a 
distance of $\sim 1.4$~kpc. 
This peculiar source  is a (possible) quadruple stellar
system that shows radio structures over different
scales, from about 10 milliarcseconds (mas) to about 30 arcseconds [3].
At the 10 mas scale, 
there is a non-thermal, arc-like structure that traces the CWR
between the wind of an eclipsing contact 
binary  and that of an unseen nearby companion. 
About $1''$ to the NE of the contact binary, there is another 
non-thermal  component that results from the interaction of the winds
of the contact binary and that of a known B-type star. Finally, there is
an extended ($\sim 30''$) synchrotron structure that could also be associated 
with Cyg~OB2~\#5. In Figure~1 (left) we show a sketch of the 
three non-thermal structures that compose this complex source.
In this study we model the $\gamma$-ray emission produced in the compact
CWR (produced by the contact binary and an unseen companion). 

\section{Particle acceleration and losses}

\begin{figure}
\includegraphics[height=.35\textheight , angle=-90]{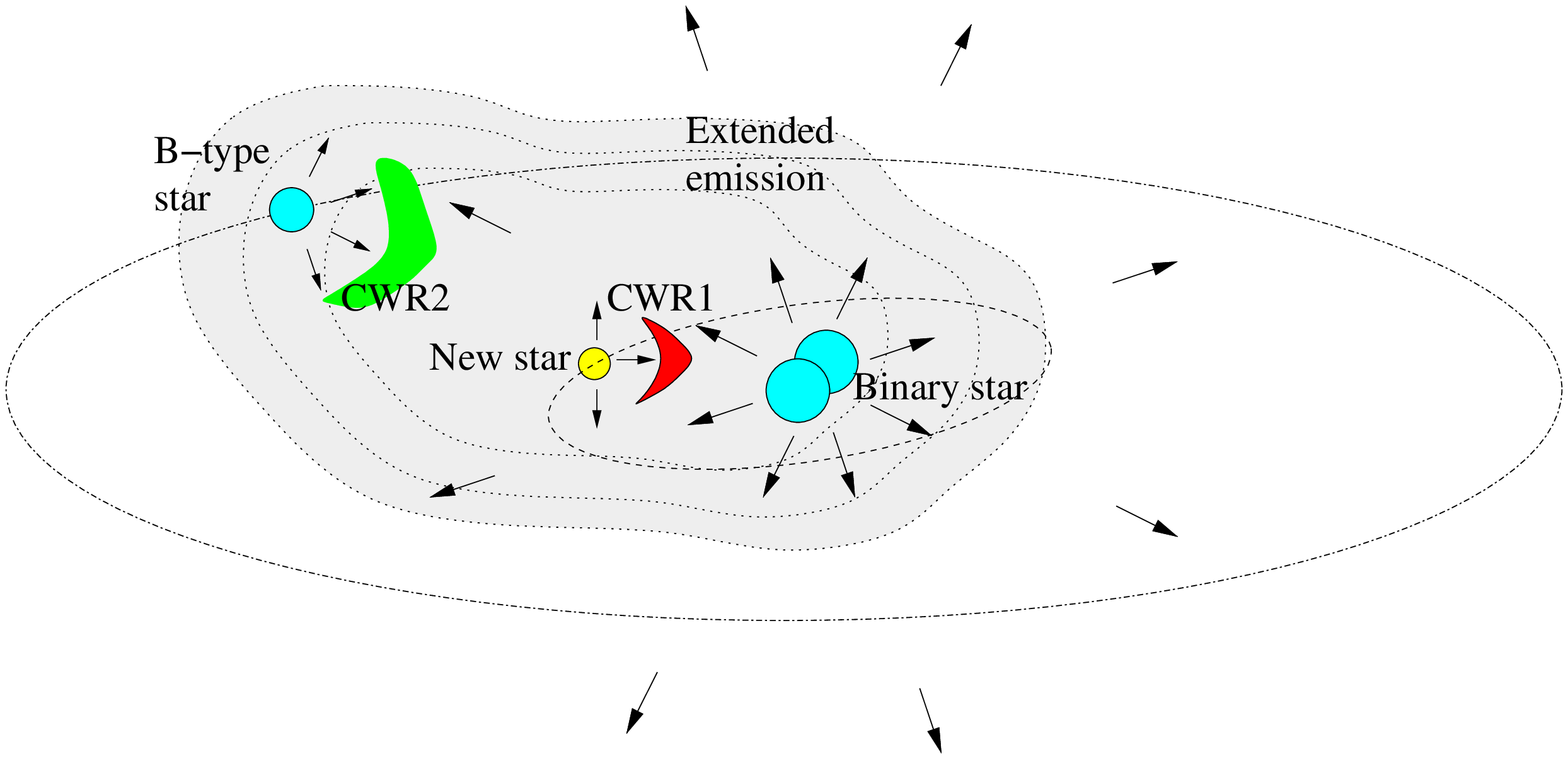}
\includegraphics[height=.35\textheight , angle=-90]{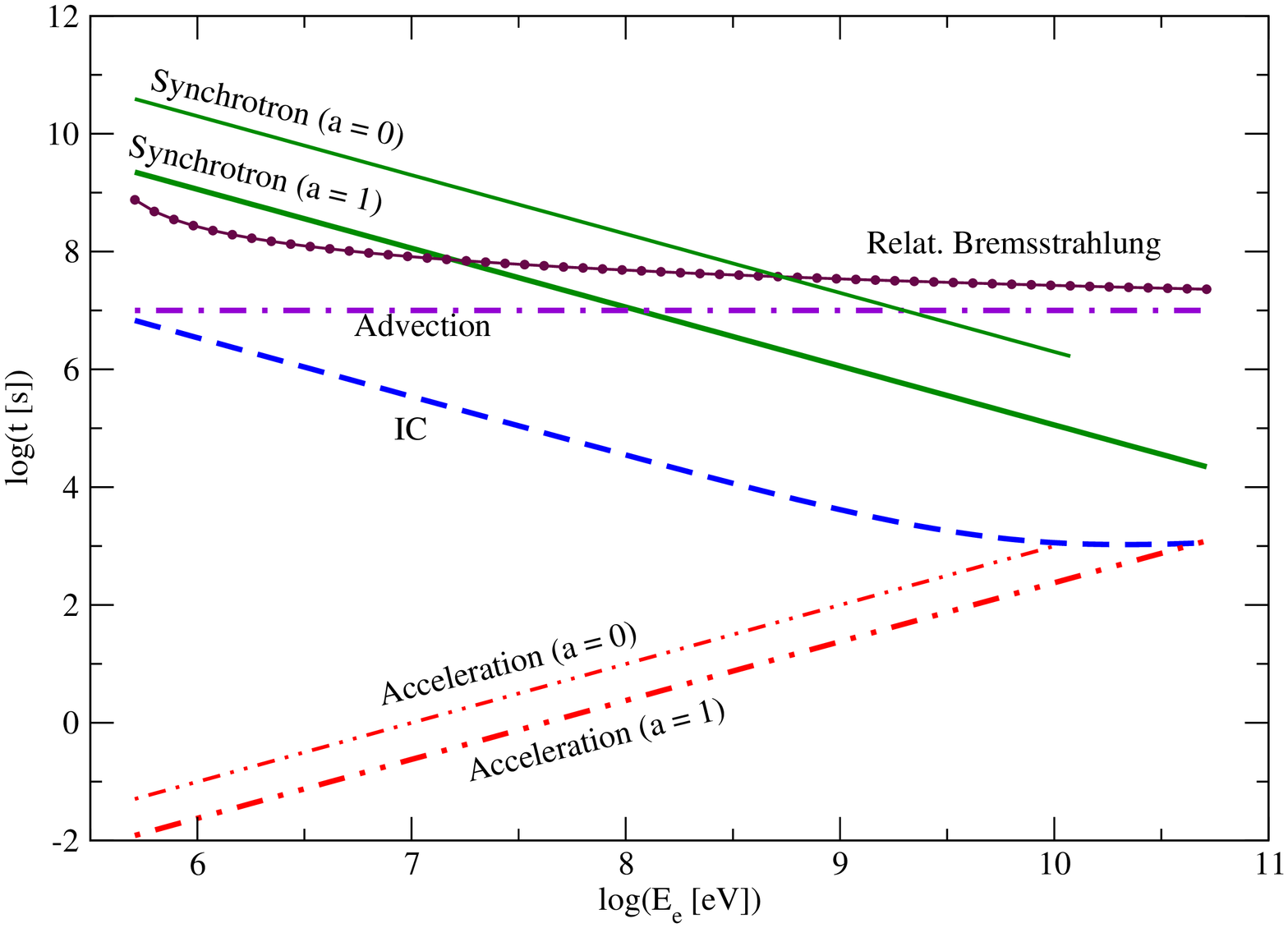}
\caption{\emph{Left}: Compact and extended components of the system 
Cyg~OB2~\#5. The sketch is not drawn to scale. \emph{Right}: 
Escape,  radiative cooling 
and acceleration times for the cases discussed in the text.}
\end{figure}

Both stars of the compact CWR are separated a distance $D=x_{\rm p} + x_{\rm s}$, 
and the CWR
is located at a distance $x_{\rm p}$ and $x_{\rm p} = \sqrt{\beta}\,x_{\rm s}$ 
from the primary and 
the secondary components of the system, respectively.
Being $\beta \equiv \dot M_{\rm p} v_{\infty,{\rm p}}/\dot M_{\rm s} v_{\infty,{\rm s}}$,
where $\dot M$ is the mass-loss rate, and $v_{\infty}$ is the terminal 
wind velocity, the stagnation point is located 
closer to the less powerful star (the unseen companion in our case).
Characterizing the wind of the contact binary by
$v_{\infty,{\rm p}} \sim 1500$~km~s$^{-1}$ and 
$\dot M_{\rm p} \sim 2\times10^{-5}$~M$_{\odot}$~yr$^{-1}$, and considering that the
companion is a B0.5 early-type star [4] $x_{\rm s}$ results  $\sim 1.5$~AU,
where we have estimated $x_{\rm p} = 25$~AU from radio data. 
At the location of the stagnation point, a discontinuity surface
and a double bow-shock structure is created. 
In long period binary systems, the formed CWR is thick
and  it is not affected by disruptive 
instabilities. Thus, CWRs are  good places for particle
acceleration, since at least one of the shocks is adiabatic and the 
shocked region will be sufficiently large to allow charged particles to reach 
relativistic energies. 
In this work we assume Fermi-I as the operating mechanism to 
accelerate particles. Being the wind of the contact binary
the most powerful, we will consider only the acceleration of 
particles in the shock produced by this wind.

The detection of synchrotron emission is an evidence that there is a 
population of relativistic electrons ($e$) in the CWR. However, protons ($p$) 
can be accelerated as well. Particles accelerated by the Fermi-I mechanism
follow a power-law energy distribution $N_{e,p} = K_{e,p} E_{e,p}^{-2.1}$,
where we have assumed that both electrons and 
protons are injected with the same spectral index $2.1$.
Considering that the same number of electrons and protons are accelerated
at energies $\leq m_ec^2$ [5], and neglecting ionization/coulomb 
losses, we relate $K_p = (m_p/m_e)^{(2.1-1)/2} K_e$.

From the measured synchrotron flux $S_{\nu}(\nu = 8.4 {\rm GHz}) = 3$~mJy [4] 
we estimate the magnetic field $B$ in the CWR
by equating magnetic  and non-thermal 
energy densities: $U_{\rm B} = U_e + a U_p$, where $U_e$ and $U_p$ are
the energy densities of relativistic electrons and protons, 
respectively,
and $a$ takes two values: $0$ (no protons are accelerated) and $1$ (protons
are accelerated as well as electrons). We obtain $B = 0.07$~G in the 
former case and $0.29$~G in the latter. 
Once we know $B$, we can calculate the acceleration and synchrotron cooling 
time. To calculate the losses by Inverse Compton 
(IC) scattering, the largest density of target photons 
($U_{\rm ph} = 4.45$~erg~cm$^{-3}$ at the location of the CWR) is provided  
by the unseen companion with luminosity 
$L_{\rm s} \sim 6.1\times10^4$~L$_{\odot}$ [4]. 
In addition to synchrotron and IC
losses,  accelerated electrons in the field of ions in the wind
can produce relativistic Bremsstrahlung radiation at $\gamma$-ray energies.
However, the density of these ions in the CWR is not enough 
($n_{\rm w} = 1.3\times10^{7}$~cm$^{-3}$) for the relativistic
Bremsstrahlung losses to be more efficient than IC cooling.
Thus, the maximum energy of accelerated electrons is constrained
by IC losses, given $E_{e,{\rm max}} \sim 40$ and $93$~GeV in the cases with 
$a=0$ and 1, respectively, as is shown in Fig.~1 (right).

Relativistic protons interacting with ions in the stellar winds
produce neutral pions  and then $\gamma$-rays. However, 
relativistic hadrons escape from the CWR on a time shorter than the 
$pp$ cooling time. At energies $E_p < 5$~TeV,
advection losses are dominant ($t_{\rm adv} \sim
4 x_{\rm p}/v_{\infty,{\rm p}} \sim 10^7$~s), 
and at larger energies the diffusion escape become
more relevant.  This produces that $N_p$ shows a break at energy 
$E_{\rm b} = 5$~TeV. 
The maximum energy of protons is constrained by diffusion losses
given $E_{p,{\rm max}} \sim 50$~TeV.

\begin{figure}
\includegraphics[height=.35\textheight , angle=-90]{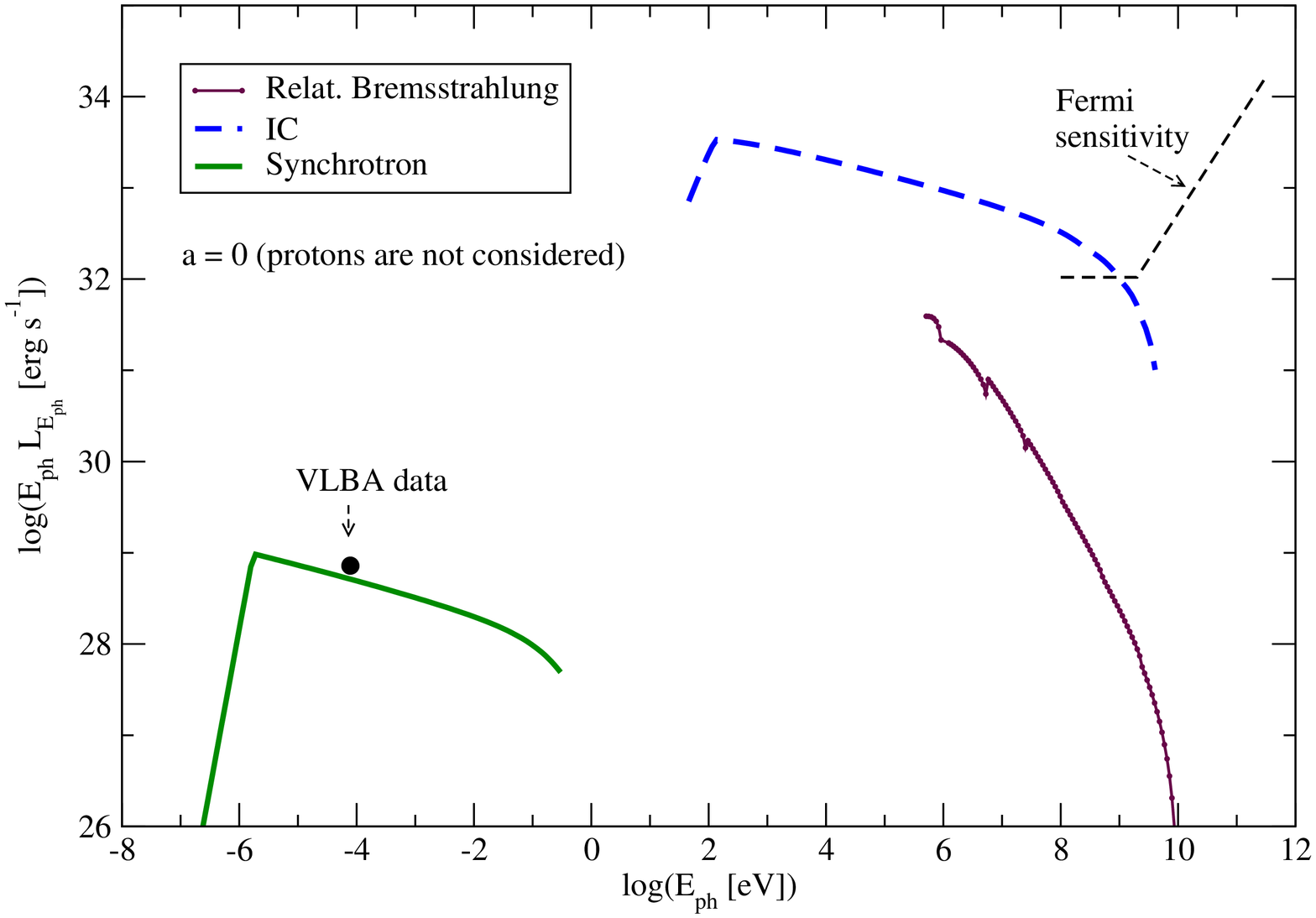}
\includegraphics[height=.35\textheight , angle=-90]{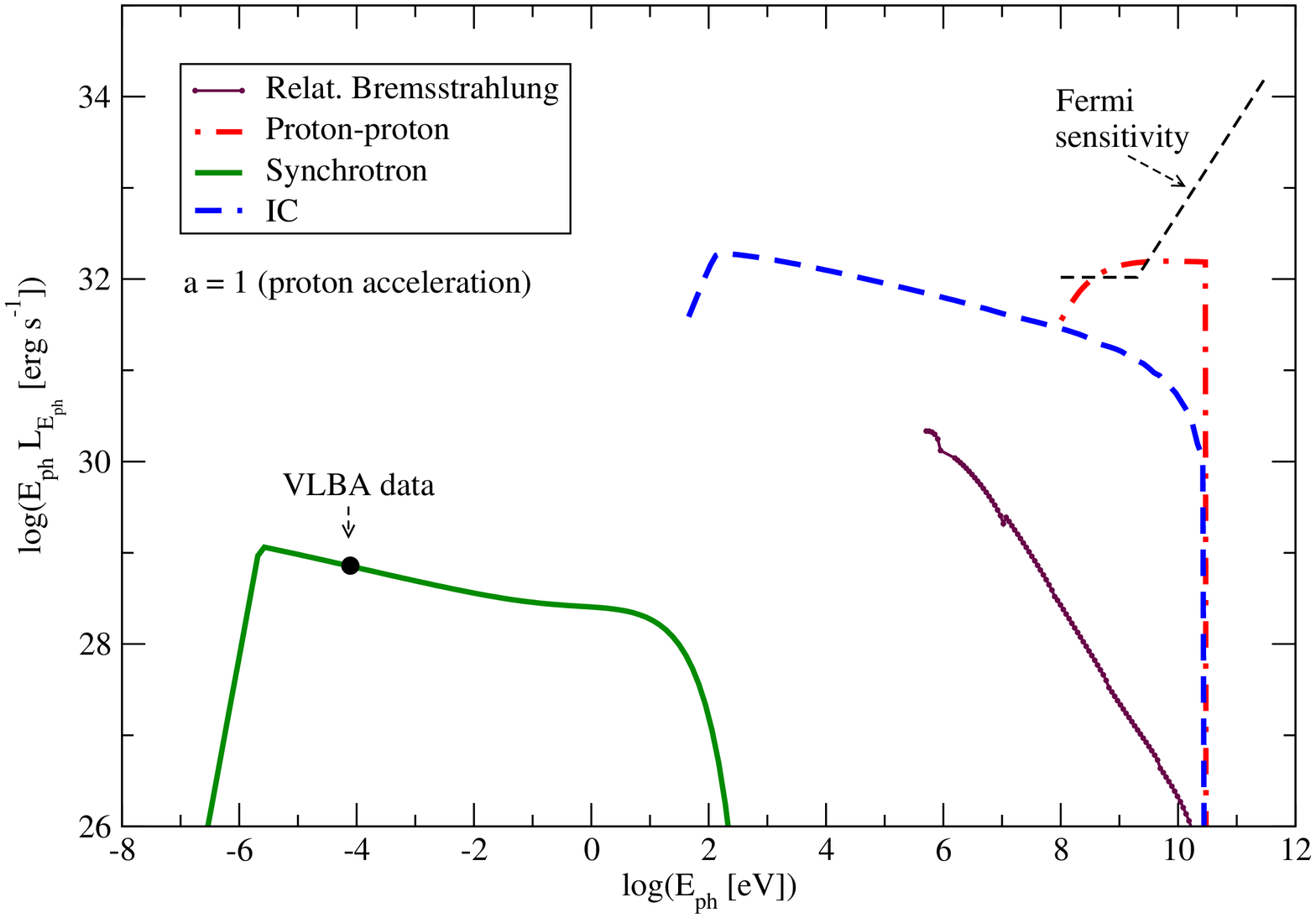}
\caption{Spectral energy distributions  for the cases 
discussed in the text. 
The \emph{Fermi} sensitivity is also plotted in order to show that the 
source is detectable by this instrument.}
\end{figure}

\section{Gamma-ray emission and discussion} 

We computed the spectral energy distribution (SED) of 
synchrotron, IC, relativistic Bremsstrahlung, and $pp$,
 in the  CWR  considering the distributions
 $N_e$ and $N_p$ derived previously 
by adjusting the synchrotron spectrum with radio data and below the
equipartition condition.
We obtain that the most important radiation mechanism to 
produce $\gamma$-rays in the compact  CWR of Cyg~OB2~\#5 is the IC scattering 
due to the dense photon stellar field provided by the unseen companion
when only electrons are considered (case with $a = 0$). In this case   
synchrotron, IC and relativistic Bremsstrahlung bolometric luminosities result
$L_{\rm syn} \sim 1.5\times10^{30}$,  $L_{\rm IC} \sim 2.6\times10^{34}$, and
$L_{\rm Brem} \sim 5.4\times10^{31}$~erg~s$^{-1}$, respectively, as
is shown in Fig.~2 (left). 
In the case with $a = 1$, the non-thermal energy is distributed between
 electrons and protons
resulting that $U_p > U_e$ as a consequence of $m_p > m_e$ [5].
The achieved bolometric  luminosities are $L_{\rm syn} \sim 1.7\times10^{30}$, 
$L_{\rm IC} \sim 1.7\times10^{33}$, and 
$L_{\rm Brem} \sim 2.9\times10^{30}$~erg~s$^{-1}$. In this case 
proton-proton collisions are the most important cooling channel 
in the $\gamma$ domain, reaching a bolometric luminosity 
$L_{pp} \sim 7.8\times10^{32}$~erg~s$^{-1}$, as
is shown in Fig.~2 (right).

The $\gamma$-ray spectrum is attenuated
at energies greater than $\sim 0.3$~TeV, where the optical depth 
due to photon-photon electron-positron pair production 
($\gamma + \gamma \rightarrow e^+ + e^-$)
in the intense stellar radiation field of the B0.5 companion star  reaches 
non-negligible value. In the case with $a=0$, photon-photon 
absorption is not relevant because $E_{\rm ph} \sim 0.3$~TeV is larger than 
the maximum  energy of photons produced there, but in the case with $a=1$ 
 the attenuation is  significant.

From Fig.~2 it can be seen that the $\gamma$-ray emission produced in the
compact CWR of
Cyg~0B2~\#5 (by the simple model described in the present contribution) can 
be detected with the \emph{Fermi} satellite. In particular, when only 
the relativistic electrons (that produce the detected synchrotron emission) 
are considered, IC scattering of photons from the unseen companion reach 
specific luminosity values in the GeV domain detectable with \emph{Fermi} in a 
deep exposure.
When  protons are also accelerated up to relativistic energies, $\gamma$-rays
from $pp$ collisions are also detectable by \emph{Fermi}. 
These results shown that 
even in the case in which the \emph{Fermi} detection from the Cyg~OB2 region 
has been associated with the pulsar PSR~J2032+4127,  
under certain conditions
a significant amount of $\gamma$-rays can be produced in Cyg~OB2~\#5. 
The detection of $\gamma$ rays from CWRs can 
help us to understand the properties of stellar winds, as well as
the development of high-energy processes in stellar environments. 


\begin{theacknowledgments}
This work is supported by CONACyT, Mexico, and PAPIIT, UNAM.
\end{theacknowledgments}


\bibliographystyle{aipproc}   

\end{document}